\documentclass[letter,twocolumn]{jpsj3}
\usepackage{times,bm}
\newcommand\hc{{\rm H.c.}}
\newcommand\ket[1]{| #1 \rangle}
\newcommand\bra[1]{\langle #1 |}

\title{Anomalous Hall Effect in Ferromagnetic Metals: \\ Role of Phonons at Finite Temperature}
\author{%
\name{Atsuo \surname{Shitade}}$^1$\thanks{present address: Department of Physics, Kyoto University, Kitashirakawa-Oiwakecho, Sakyo-ku, Kyoto 606-8502}%
and \name{Naoto \surname{Nagaosa}}$^{1,2,3}$}
\inst{%
\address{$^1$Department of Applied Physics, The University of Tokyo, Hongo, Bunkyo-ku, Tokyo 113-8656} \\
\address{$^2$Correlated Electron Research Group (CERG), RIKEN Advanced Science Institute (ASI), Wako 351-0198} \\
\address{$^3$Cross-Correlated Materials Research Group (CMRG), RIKEN Advanced Science Institute (ASI), Wako 351-0198}%
}
\abst{%
The anomalous Hall effect in a multiband tight-binding model is numerically studied
taking into account both elastic scattering by disorder and inelastic scattering by the electron-phonon interaction.
The Hall conductivity is obtained as a function of
temperature $T$, inelastic scattering rate $\gamma$, chemical potential $\mu$, and impurity concentration $x_{\rm imp}$.
We find that the new scaling law holds over a wide range of these parameters;
$-\sigma_{xy}= (\alpha \sigma_{xx0}^{-1} + \beta \sigma_{xx0}^{-2}) \sigma_{xx}^2 + b$,
with $\sigma_{\mu \nu}$ ($\sigma_{\mu \nu 0}$) being  the conductivity tensor (with only elastic scattering), which corresponds to the recent
experimental observation~[Phys. Rev. Lett. {\bf 103} (2009) 087206].
The condition of this scaling is examined.
Also, it is found that the intrinsic mechanism depends on temperature under a resonance condition.%
}
\kword{anomalous Hall effect, inelastic scattering, phonon}
\begin{document}
\maketitle
The anomalous Hall effect (AHE) is a phenomenon where the transverse conductivity arises
owing to the relativistic spin-orbit interaction in the presence of the spontaneous magnetization~\cite{RevModPhys.82.1539}.
This effect has attracted much attention because it is closely relevant to many fundamental questions on quantum transport in solids.
Historically, the first theory of AHE was proposed by Karplus and Luttinger (KL)
who considered the matrix elements of a current operator between different bands
and discovered the so-called anomalous velocity~\cite{PhysRev.95.1154}.
In their original paper, they did not mention the disorder effect
and were criticized by Smit who stressed that the accelerating electric field is balanced
by that of the impurity potential in the steady state~\cite{Smit1955877,Smit195839}.
He instead proposed the skew scattering mechanism,
where the transition probability for the scattering ${\bm k} \to {\bm k}^{\prime}$ is different from that for ${\bm k}^{\prime} \to {\bm k}$.
Namely, the detailed balance is violated in a scattering process, leading to the transverse current to the electric field.
Later, Berger proposed another mechanism called the side jump,
where a transverse shift in the electron trajectory occurs during the scattering~\cite{PhysRevB.2.4559,PhysRevB.5.1862}.
These two are collectively dubbed the {\it extrinsic mechanisms}, while that of KL the {\it intrinsic mechanism}.
For a long time, it has been considered that the extrinsic mechanisms are dominant,
while the intrinsic mechanism is negligible~\cite{RevModPhys.82.1539}.
Experimentally, the Hall resistivity $\rho_{xy}^{\rm tot}$ is analyzed using the expression
\begin{equation}
  \rho_{xy}^{\rm tot} = R_0 H + \rho_{xy},
\end{equation}
where $R_0$ is the ordinary Hall coefficient,
while the second term $\rho_{xy}$ is the anomalous contribution due to the spontaneous magnetization $M_{\rm s}$ and
is often empirically written as $4 \pi R_{\rm s} M_{\rm s}$ with the anomalous Hall coefficient $R_{\rm s}$. 
Furthermore, the resistivity ($\rho_{xx}$) dependence of $\rho_{xy}$ has been analyzed using
\begin{equation}
  \rho_{xy} = \alpha \rho_{xx} + \beta \rho_{xx}^2,
\end{equation}
where the first term corresponds to the skew scattering, while the second one coresponds to the side jump.
However, note that the intrinsic contribution could also contribute to the second term.

Recent development has been achieved by theoretical reformulation based on the Berry curvature in the momentum space~\cite{RevModPhys.82.1959}
and by first-principles band calculations~\cite{Fang03102003,PhysRevLett.92.037204,PhysRevB.74.195118,PhysRevB.76.195109},
which are combined to reveal the topological nature of the AHE~\cite{RevModPhys.82.1539}.
The disorder effect on the AHE has been theoretically studied to reveal the scaling behavior~\cite{PhysRevB.77.165103,PhysRevB.79.195129},
which is consistent with the systematic experimental results~\cite{RevModPhys.82.1539,PhysRevLett.99.086602}.
However, most theoretical analyses have been limited to the zero temperature limit where inelastic scattering is neglected.
The AHE at finite temperature has already been studied~\cite{Irkhin1962,PTP.27.772,PhysRev.160.421}, but remains controversial.
Therefore, a more systematic study including inelastic scattering at finite temperature is highly desired. 

Experimentally, it has been found that the skew scattering contribution is rapidly suppressed as temperature increases,
and the Hall conductivity $\sigma_{xy}$ reaches a rather steady value~\cite{RevModPhys.82.1539}.
More explicitly, an empirical formula for $\sigma_{xy}$ has been proposed by Tian {\it et al.}~\cite{PhysRevLett.103.087206} as
\begin{align}
  & -\sigma_{xy}(T, d) 
  = \rho_{xy}^{\rm ext}(d) \sigma_{xx}^2(T, d) + b,
  \label{eq:Jin1} \\
  & \rho_{xy}^{\rm ext}(d)
  = \alpha(d) \sigma_{xx}^{-1}(T \to 0, d) + \beta(d) \sigma_{xx}^{-2}(T \to 0, d),
  \label{eq:Jin2}
\end{align}
where $\sigma_{xx}(T \to 0, d)$ is the residual conductivity at the lowest temperature experimentally accessible.
They showed that this expression fits well the data in Fe thin films
with varying temperature $T$ and the thickness $d$.
The remarkable observation in ref.~\citen{PhysRevLett.103.087206} is that $b$ is independent of both $T$ and $d$,
which is identified as $-\sigma_{xy}^{\rm int}$.
Note, however, that the temperature dependence of $b$ was reported for Ni thin films~\cite{PhysRevB.85.220403}.
The ideas underlying eq.~\eqref{eq:Jin1} are that the roles of elastic and inelastic scatterings are different,
and that only the former contributes to the extrinsic mechanisms,
while the latter rapidly suppresses them.
On the other hand, the intrinsic contribution $\sigma_{xy}^{\rm int}$ is expected to be much more robust.
Although eqs.~\eqref{eq:Jin1} and \eqref{eq:Jin2} seem to work well experimentally, their justification and theoretical explanation are still lacking.

In this paper, we study both the intrinsic and extrinsic mechanisms in a simple model
taking into account both elastic and inelastic scatterings at finite temperature $T$ much lower than the Curie temperature $T_{\rm C}$.
To this end, we construct a fully spin-polarized multiband tight-binding model on the square lattice:
\begin{align}
  H
  = & -t_0 \sum_{\langle ij \rangle} c_i^{\dagger} c_j
  + \epsilon_1 \sum_i^{\rm random} s_i^{\dagger} s_i - V_1 \sum_i^{\rm random} c_i^{\dagger} s_i + \hc \notag \\
  & + \epsilon_2 \sum_i^{\rm random} p_i^{\dagger} p_i - V_2 \sum_{\langle ij \rangle}^{\rm random} e^{-i \theta_{ij}} c_i^{\dagger} p_j + \hc
  \label{eq:Anderson}
\end{align}
Conduction electrons are described by the creation and annihilation operators $c_i^{\dagger} (c_i)$.
Each impurity has the $s$ and $p^x - i p^y$ orbitals, which are described by $s_i^{\dagger} (s_i)$ and $p_i^{\dagger} (p_i)$, respectively.
The phase factor $e^{-i \theta_{ij}}$ originates from the spin-orbit-coupled $p^x- i p^y$ orbital,
where $\theta_{ij}$ is the angle from the impurity at the $j$th site to the electron at the $i$th site measured from the $x$ axis.
For simplicity, the transfer integrals between impurities are neglected.
Also, the impurity concentration $x_{\rm imp}$ is changed from zero to unity.
Therefore, this model connects two different limits.
One is the periodic three-band model in the dense limit $x_{\rm imp} \simeq 1$,
which exhibits the intrinsic mechanism due to the existence of the $p^x - i p^y$ orbital~\cite{RevModPhys.82.1539}.
The other is the single-impurity model in the dilute limit $x_{\rm imp} \ll 1$,
which exhibits the skew scattering due to the interference between the $s$ and $p$ orbitals.
Generally, two orbitals with the azimuthal quantum numbers $l$ and $l + 1$
are necessary for the skew scattering~\cite{PhysRevLett.28.303,0305-4608-3-12-014,PhysRevB.13.397,Fert1981231}.
Below, we choose $t_0 = 1$, $-\epsilon_1 = \epsilon_2 = 1$, and $V_1 = V_2 = 0.1$,
which leads to a reasonable Hall angle on the order of $10^{-3}$.

Therefore, our model is the minimal one containing all the essential features of the AHE at finite temperature.
We will consider only the electron-phonon interaction as the source of inelastic scattering, and neglect the electron-magnon interaction.
This is justified when $T, \Theta_{\rm D} \ll T_{\rm C}$ with the Debye temperature $\Theta_{\rm D}$. 
This condition is satisfied for Fe (where the new scaling laws eqs.~\eqref{eq:Jin1} and \eqref{eq:Jin2} have been proposed~\cite{PhysRevLett.103.087206})
and Co with $T_{\rm C} > 1000 {\rm K}$, but is marginal for Ni with $T_{\rm C} = 627 {\rm K}$, and is not justified for materials with lower $T_{\rm C}$.

The conductivity tensor is numerically calculated by the Kubo formula~\cite{JPSJ.12.570},
\begin{equation}
  \sigma_{\mu \nu}
  = \frac{2 \pi}{i N_{\rm s}} \sum_{mn} \frac{f(\xi_m) - f(\xi_n)}{\xi_m - \xi_n}
  \frac{\bra{m}J^{\mu}\ket{n} \bra{n}J^{\nu}\ket{m}}{\xi_m - \xi_n + i \gamma},
  \label{eq:kubo}
\end{equation}
for each impurity configuration.
A set of conductivities are averaged over $9600$ configurations for $0 < x_{\rm imp} \leq 0.3$,
$6400$ for $0.4 \leq x_{\rm imp} \leq 0.6$, and $3200$ for $0.7 \leq x_{\rm imp} < 1$.
Here $f(\xi) = (e^{\xi/T} + 1)^{-1}$ is the Fermi distribution function
with the energy $\xi$ measured from the chemical potential $\mu$ and finite temperature $T$.
The inelastic scattering rate $\gamma$ is phenomenologically introduced as the imaginary part of the self-energy,
which in reality depends on temperature and frequency.
The effect of the frequency dependence of the self-energy will be discussed later.
The factor $2 \pi$ is due to the conductivity unit of $e^2/h = (2 \pi)^{-1}$,
and $N_{\rm s} = 30 \times 30$ is the number of sites.
The eigenstate $\ket{n}$ is represented on the real-space basis
because we consider the general impurity concentration $x_{\rm imp}$.
Our numerical calculations can fully take into account the impurity effects for a given impurity configuration
because we employ the numerical diagonalization and the Kubo formula itself is exact.
The {\it ab initio} Korringa-Kohn-Rostoker method combined with the coherent potential approximation was employed
to investigate the AHE~\cite{PhysRevLett.105.266604},
but it is applicable only to elastic scattering and zero temperature.
On the other hand, our method can reveal a unified picture of the AHE at finite temperature with the inelastic scattering rate $\gamma$.

\begin{figure}
  \centering
  \includegraphics[clip,width=0.48\textwidth]{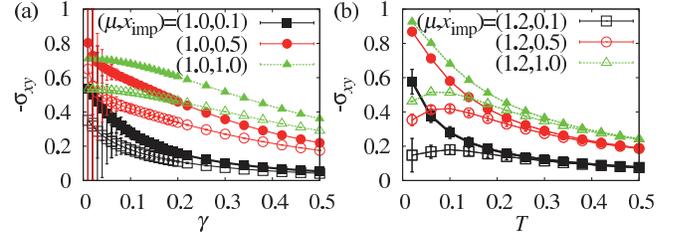}
  \caption{%
  (Color online) 
  Calculated Hall conductivity $-\sigma_{xy}$
  (a) as a function of inelastic scattering strength $\gamma$ for $T = 0.1$
  and (b) as a function of temperature $T$ for $\gamma = 0.1$.
  $x_{\rm imp} = 0.1, 0.5, 1.0$ and $\mu = \epsilon_2 = 1, 1.2$.%
  }
  \label{fig1}
\end{figure}

\begin{figure}
  \centering
  \includegraphics[clip,width=0.48\textwidth]{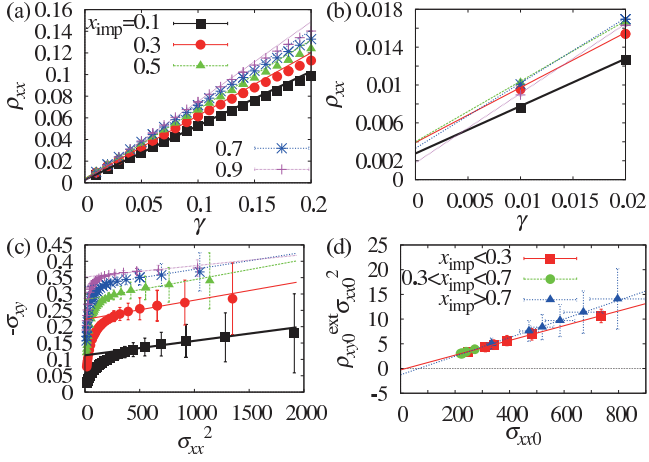}
  \caption{%
  (Color online)
  (a) Longitudinal resistivity $\rho_{xx}$ as a function of $\gamma$ and (b) its magnified image
  for $x_{\rm imp} = 0.1, 0.3, 0.5, 0.7, 0.9$.
  (c) Hall conductivity $-\sigma_{xy}$ as a function of $\sigma_{xx}^2$ by changing $\gamma$, which is based on eq.~\eqref{eq:fit1b}.
  (d) Plot for $\rho_{xy0}^{\rm ext} \sigma_{xx0}^2$ vs $\sigma_{xx0}$ obtained using eq.~\eqref{eq:fit2}.
  $T = 0.3$ and $\mu = \epsilon_2 = 1$.
  }
  \label{fig2}
\end{figure}

\begin{figure}
  \centering
  \includegraphics[clip,width=0.48\textwidth]{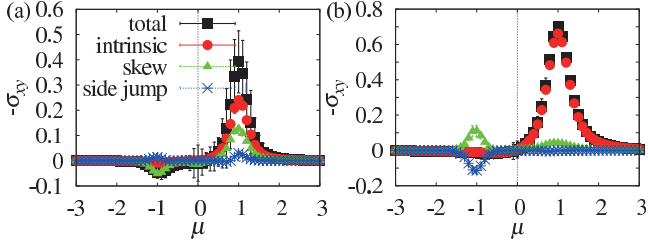}
  \caption{%
  (Color online)
  Intrinsic (red filled circle), skew scattering (green open triangle), and side jump (blue star) contributions
  extracted from the total Hall conductivity (black filled square) as functions of the chemical potential $\mu$.
  $T = 0.1$, $\gamma = 0.05$, and (a) $x_{\rm imp} = 0.1$ and (b) $x_{\rm imp} = 0.9$.%
  }
  \label{fig3}
\end{figure}

We can obtain the Hall conductivity as a function of $T$, $\gamma$, $\mu$, and $x_{\rm imp}$.
In Fig.~\ref{fig1}(a), we can find the strong $\gamma$ dependence of $-\sigma_{xy}$ in the dilute regime,
and the weak $\gamma$ dependence in the dense regime.
The temperature dependence at $\mu = \epsilon_2$ is strong, but is weak in other cases, as seen in Fig.~\ref{fig1}(b).
Thus, the calculated Hall conductivity shows different dependences on many parameters, and is analyzed below.
First, we can find the $\gamma$ dependence of $\sigma_{xx}$ in Figs.~\ref{fig2}(a) and \ref{fig2}(b),
which is fitted by $\sigma_{xx}^{-1}(T, \gamma) = \sigma_{xx0}^{-1}(T) (\gamma + \gamma_0)/\gamma_0$,
i.e., the Matthiessen's rule holds.
Thus, $\gamma_0$ obtained by fitting can be assigned to the elastic scattering rate due to disorder,
while $\gamma$ is assigned to the inelastic scattering rate.
As shown in Fig.~\ref{fig2}(c), $-\sigma_{xy}$ is a linear function of $\sigma_{xx}^2$ by changing $\gamma$
in the dilute, intermediate, and dense regions.
Namely, the relation 
\begin{equation}
  -\sigma_{xy}(T,\gamma) = \rho_{xy0}^{\rm ext}(T) \sigma_{xx}^2(T,\gamma) + b(T)
  \label{eq:fit1b}
\end{equation}
holds for any impurity concentration $0 < x_{\rm imp} < 1$.
Here, the subscript $0$ does not mean the $T \to 0$ limit, but the $\gamma \to 0$ limit.
The temperature dependences of $\rho_{xy0}^{\rm ext}(T)$ and $b(T)$ come from the Fermi distribution function only.
In real experiments, the temperature dependence of $-\sigma_{xy}(T,\gamma)$ mainly comes from that of $\gamma = \gamma(T)$.
Therefore,  eq.~\eqref{eq:fit1b} almost corresponds to eq.~\eqref{eq:Jin1} experimentally proposed in ref.~\citen{PhysRevLett.103.087206}.
By combining eq.~\eqref{eq:fit1b} and the Matthiessen's rule,
we can conclude that the intrinsic contribution $-\sigma_{xy}^{\rm int} = b(T)$ is robust against inelastic scattering,
while the extrinsic contribution is rapidly suppressed as $-\sigma_{xy}^{\rm ext} = \rho_{xy0}^{\rm ext} \sigma_{xx0}^2/(\gamma/\gamma_0 + 1)^2$.
The extrinsic contribution in the first term can be further separated into the skew scattering and side jump contributions using
\begin{equation}
  \rho_{xy0}^{\rm ext}(T)
  = \alpha(T) \sigma_{xx0}^{-1}(T) + \beta(T) \sigma_{xx0}^{-2}(T),
  \label{eq:fit2}
\end{equation}
corresponding to the second scaling law eq.~\eqref{eq:Jin2}.
As shown in Fig.~\ref{fig2}(d),
$\rho_{xy0}^{\rm ext} \sigma_{xx0}^2$ is a linear function of $\sigma_{xx0}$ in the dilute and dense regions, respectively,
but not in the intermediate region
because higher-order perturbations with respect to disorder potential are relevant and the band structure is ill-defined.
In Fig.~\ref{fig3}, the separated contributions of the intrinsic mechanism, skew scattering, and side jump
are plotted as functions of the chemical potential.
In Fig.~\ref{fig3}(a) for $x_{\rm imp} = 0.1$ in the dilute limit,
the skew scattering contribution $-\sigma_{xy}^{\rm skew} = \alpha(T) \sigma_{xx}^2(T, \gamma)/\sigma_{xx0}(T)$ is almost dominant
at $\mu \simeq \epsilon_1$ and $\epsilon_2$.
This is consistent with the fact that the skew scattering is from the interference between the $s$ and $p$ orbitals.
On the other hand, in Fig.~\ref{fig3}(b) for $x_{\rm imp} = 0.9$ in the dense limit,
the intrinsic mechanism is dominant.
Note that the $s$ orbital of an impurity does not contribute to the Hall conductivity in this limit.

\begin{figure}
  \centering
  \includegraphics[clip,width=0.48\textwidth]{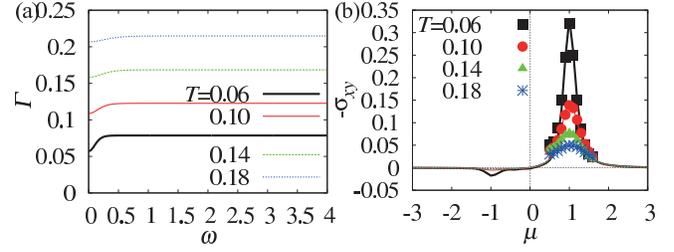}
  \caption{%
  (Color online) 
  (a) Frequency dependence of the imaginary part of the self-energy, $\Gamma(\omega) = -\Im \Sigma(\omega)$.
  (b) Chemical potential dependences of the Hall conductivities calculated using eqs.~\eqref{eq:kubo2} (points) and \eqref{eq:kubo} (lines).
  $x_{\rm imp} = 0.1$ and $\Theta_{\rm D} = 0.1$.%
  }
  \label{fig4}
\end{figure}

Now, we examine the validity of eq.~\eqref{eq:kubo} with the constant $\gamma$ in the above calculation. 
The imaginary part of the self-energy $\Gamma(\omega) = -\Im \Sigma(\omega)$ depends on frequency, and the vertex correction exists.
According to the Migdal's theorem~\cite{migdal1958},
the vertex correction is on the order of $O(\Theta_{\rm D}/E_{\rm F}) = 10^{-3}$ and negligible. 
As for the self-energy, we have calculated the lowest order as shown in Fig.~\ref{fig4}(a)~\cite{T1964410}.
It is found that the frequency dependence of $\Gamma(\omega)$ is appreciable there only when $\omega, T < \Theta_{\rm D}$; it is negligible otherwise. 
This frequency dependence is taken into account using
\begin{align}
  \sigma_{\mu \nu}
  = & \frac{2 \pi}{i N_{\rm s}} \sum_{m n} \bra{m} J^{\mu} \ket{n} \bra{n} J^{\nu} \ket{m} \notag \\
  & \times \int {\rm d} \omega A_m(\omega) \int {\rm d} \omega^{\prime} A_n(\omega^{\prime})
  \frac{f(\omega) - f(\omega^{\prime})}{(\omega - \omega^{\prime} + i \delta)^2},
  \label{eq:kubo2}
\end{align}
where $A_m(\omega) = \pi^{-1} \Gamma(\omega)/((\omega - \xi_m)^2 + \Gamma^2(\omega))$ is the spectral function
and $\delta$ is infinitesimal.

Since the effect of inelastic scattering is mainly on the extrinsic mechanisms,
we choose the case of $x_{\rm imp}=0.1$ where the skew scattering has a large contribution.
Figure~\ref{fig4}(b) shows the Hall conductivities calculated using eqs.~\eqref{eq:kubo2} and \eqref{eq:kubo}.
The former is found to be approximated by the latter with
$\gamma/2 = \Gamma(\omega = 0)$
rather than $\Gamma(\omega \to \infty)$ down to $T = 0.6 \Theta_{\rm D} = 0.06$,
which is the lowest temperature at which we could obtain sufficiently accurate data of numerical calculation.
This suggests that the Fermi surface term, rather than the Fermi sea term, 
is dominant in the extrinsic mechanisms similarly to the longitudinal conductivity.
At lower temperatures, when $\Gamma(\omega = 0) \propto \Theta_{\rm D} (T/\Theta_{\rm D})^3$
is less than $\Gamma(\omega = \Delta) \propto \Theta_{\rm D} (\Delta/\Theta_{\rm D})^3$
with $\Delta \simeq V^2/E_{\rm F} \simeq 10^{-2}$ being the small anticrossing induced by the spin-orbit interaction,
the Fermi sea term becomes dominant and the above approximation breaks down.
As for the longitudinal $\sigma_{xx}$, the frequency region 
$\omega \sim T$ is relevant and hence $\gamma/2=\Gamma(\omega=0)$ is also justified.
Thus, the scaling relation confirmed above is expected to hold for $T > \Delta$
even when the structure of the self-energy is taken into account.
On the other hand, inelastic scattering by magnons is qualitatively different from that by phonons
since the electron-magnon interaction involves a spin flip.
It was shown that the intrinsic contribution is canceled by the side jump contribution
and that the skew scattering is forbidden in the quasielastic region, i.e., near the Curie temperature~\cite{PhysRevB.83.125122}.

\begin{figure} 
  \centering
  \includegraphics[clip,width=0.48\textwidth]{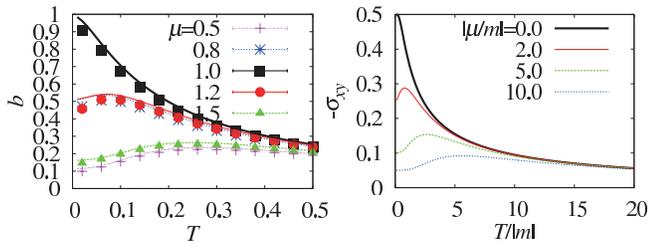}
  \caption{%
  (Color online)
  (a) Temperature dependence of the intrinsic contribution $-\sigma_{xy}^{\rm int} = b(T)$
  at  $\mu \simeq \epsilon_2 = 1$ for $x_{\rm imp} = 0.98$.
  Each line is obtained by direct calculations of $-\sigma_{xy}^{\rm int}$ on the momentum-space basis
  for $x_{\rm imp} = 1$ and $\gamma = 0.01$.
  (b) Temperature dependence of the Hall conductivity for the massive Dirac Hamiltonian eq.~\eqref{eq:Dirac}.%
  }
  \label{fig5}
\end{figure}

On the other hand, the intrinsic mechanism is robust against inelastic scattering,
and its temperature dependence is determined only by the Fermi distribution function.
Figure~\ref{fig5}(a) shows that $b(T)$ depends on temperature only under the resonance condition $\mu \simeq \epsilon_2$,
while it is almost independent in the case of off resonance.
Actually, in the massive Dirac Hamiltonian written as
\begin{equation}
  H_{\bm k} = k^y \sigma^x - k^x \sigma^y + m \sigma^z,
  \label{eq:Dirac}
\end{equation}
where $\sigma^x$, $\sigma^y$, and $\sigma^z$ are the Pauli matrices,
the temperature dependence of the Hall conductivity is strong and monotonic when the chemical potential is in the gap,
but becomes weak and nonmonotonic away from the gap, as shown in Fig.~\ref{fig5}(b).
The temperature dependence of the Hall conductivity in ferromagnetic Ni films observed in ref.~\citen{PhysRevB.85.220403}
can be explained by the intrinsic mechanism.

Finally, let us discuss the relevance of our theory to experiments.
In ref.~\citen{PhysRevLett.103.087206}, the $b$ term in eq.~\eqref{eq:Jin1} is responsible for the intrinsic mechanism
because it is independent of the film thickness $d$.
Although we have no parameter corresponding to $d$,
$b(T)$ at $x_{\rm imp} = 0.98$ is well fitted by $-\sigma_{xy}$ at $x_{\rm imp} = 1$ on the momentum-space basis in Fig.~\ref{fig5}(a),
which validates the identification of $b(T)$ as the intrinsic mechanism.
Therefore, our calculations are consistent with the scaling plot in ref.~\citen{PhysRevLett.103.087206} in the dilute and dense limits.
However, in the case of alloys, i.e., intermediate $x_{\rm imp}$,
the scaling law eq.~\eqref{eq:fit1b} holds, but not eq.~\eqref{eq:fit2}, as indicated by Figs.~\ref{fig2}(c) and \ref{fig2}(d), respectively.
Recently, the effects of inelastic scattering on the AHE have been
investigated by measuring the Lorenz ratio $L_{xy} = \kappa_{xy}/\sigma_{xy} T$
~\cite{PhysRevLett.100.016601,PhysRevB.79.100404,PhysRevB.81.054414,RevModPhys.82.1539},
where $\kappa_{xy}$ is the thermal Hall conductivity.
The suppression of the anomalous part of the Lorenz ratio $L_{xy}^{\rm A}$ from the canonical value $L_0 = (\pi k_{\rm B}/e)^2/3$
indicates the effect of inelastic scattering.
It was observed that $L_{xy}^{\rm A}$ is suppressed up to $90 {\rm K}$ in disordered Fe where the skew scattering contribution is dominant,
while it is almost constant in disordered Ni where the intrinsic contribution is dominant.
This means that the extrinsic mechanisms are suppressed by inelastic scattering, but not the intrinsic mechanism,
which corresponds qualitatively well to our finding.

In conclusion, we have numerically studied the effects of elastic and inelastic scatterings on the AHE
at finite temperature much lower than the Curie temperature $T_{\rm C}$.
The scaling relation eq.~\eqref{eq:fit1b},
which states that the extrinsic mechanisms are rapidly suppressed by inelastic scattering, while the intrinsic mechanism is robust,
was found to hold for any impurity concentration,
as well as eq.~\eqref{eq:fit2} for separating the skew scattering and the side jump in the dilute and dense regimes.
These two equations correspond to the empirical scaling relations that have recently been proposed in Fe films~\cite{PhysRevLett.103.087206}.
Even when the frequency dependence of the self-energy due to the electron-phonon interaction is seriously taken into account,
the Fermi surface contribution at $\omega = 0$ is dominant in the extrinsic mechanisms;
hence the scaling laws hold at higher temperature than the small anticrossing $\Delta$.
The intrinsic mechanism depends on temperature when the resonance condition is satisfied,
corresponding to the observed temperature dependence of the intrinsic mechanism in Ni films~\cite{PhysRevB.85.220403}.

A. S. was supported by Grant-in-Aid for the Japan Society for the Promotion of Science (JSPS) Fellows.
This work was supported by MEXT Grant-in-Aid No. 24244054,
Strategic International Cooperative Program (Joint Research Type) from the Japan Science and Technology Agency,
and JSPS through its ``Funding Program for World-Leading Innovative R{\&}D on Science and Technology (FIRST Program).''

\end{document}